Gebrehiwet Gebrekrstos Lema*

# Free space optics communication system design using iterative optimization

**Abstract:** Free Space Optics (FSO) communication provides attractive bandwidth enhancement with unlicensed bands worldwide spectrum. However, the link capacity and availability are the major concern in the different atmospheric conditions. The reliability of the link is highly dependent on weather conditions that attenuate the signal strength. Hence, this study focuses to mitigate the weather and geographic effects using iterative optimization on FSO communication. The optimization maximizes the visibility distance while guaranteeing the reliability by minimizing the Bit Error Rate (BER). The wireless optical communication system is designed for the data rate of 10 Gbps. The performance of the proposed wireless optical communication is compared against the literature in terms of visibility distance, quality factor, BER, and Eye diagram at different atmospheric conditions. The simulation results have shown that the proposed work has achieved better performance.

## 1 Introduction

The wireless communication has shown pragmatic development. This increased customer attraction has led to significant demand for high Quality of Service (QoS). Though optical communication has been providing tremendous data rates in a glass guided communication link, the benefits of Free Space Optics (FSO) was mot exploited even though it has significant data rate, security, and reliability benefits over the ordinary RF wireless communications. The optical communication is not accessible in remote areas because of both the deployment difficulty and the cost-ineffective. Recently, FSO communication [1–3] has shown an attractive alternative solution that replaces the radio and microwave communication with Gigabits data rate. FSO provides local area network unlicensed spectrum, simple deployment, free electromagnetic signal interference, and extremely high data rate [4]. However, these significant ranges of FSO benefits are challenged by its high susceptibility to attenuation because of the weather and turbulence conditions [5]. The light beam loss happens because of the absorption due to molecular diffusion and scattering caused by fog, rain, snow, and haze [6]. The atmospheric turbulence happens because of the scattering, absorption, and dispersion due to fog, haze, mist, snow, and rain.

The need for high-speed Internet is significantly growing with the fast expansion of smartphones. The customers can use it on different online applications, audio/video streaming, videoconferencing, online messaging, and web browsing [7, 8]. To estimate the average usage by these applications, the average number of passengers on one train ranges from 500 to 1300 [9] which requires several Gbps data rates. For example, a resolution of 1280 × 720 pixels YouTube video user requires 2500 Kbps data rates [10]. The collective 500 users video demand requires a 1.25 Gbps data rate. However, imagine how this data rate demand is not easy to achieve using the usual RF communications because of the Doppler effect due to movement, frequent handovers, and operational frequencies and bandwidths [11]. In general, the FSO connections are becoming a fascinating alternative for copper, RF, and fiber optic communication techniques, in terms of speed, costing, distance, and mobility.

Recently, 2018 [12], An adaptive beam has proposed that adapts its divergence angle according to the receiver aperture diameter and its communication distance to improve the received power. However, neither the Bit Error Rate (BER) reduction nor the visibility distance enhancement was significant enough. For different atmospheric turbulences, the digital modulations including amplitude shift keying and pulse position modulation techniques [13] are evaluated. However, the data rate was limited to 2.5 Gbps and the visibility distance was limited. Besides, there was no adaptive

---

*Corresponding author: Gebrehiwet Gebrekrstos Lema, School of Electrical and Computer Engineering, Mekelle University, Mekelle, Ethiopia, E-mail: g.jcool.com@gmail.com. https://orcid.org/0000-0001-5703-1391

concept that enhances the visibility distance according to the atmospheric conditions.

The overall wireless optical communication has tremendous benefits over the RF communication as it has a higher operating frequency and hence better data rate [14], however, the atmospheric condition prone problems are challenging. In 2017 [15], attractive data rate (10 Gbps) has been achieved at different optical bands, however, the visibility distance was limited to only 500 m.

[16] evaluates three optical transmission windows performance on bad weather conditions, however, it doesn't propose any atmospheric turbulence mitigating technique. In 2018 [17], again other types of digital modulations (namely amplitude shift keying and phase shift keying) were compared and the latter has shown better performance. However, it has limited increment and it is insufficient for the breaking multimedia requirement of the generation and beyond.

Quite recently 2019 [18], a transmit power adapting transmitter and receiver design was used to combat the atmospheric problems, how8ever, the design applies expensive parameters to overcome the channel impairments. Increasing the transmit power contradicts the energy efficiency of the current and future cellular communications. The former research, [19], has also shown the significance and challenges of the transmit power for FSO. More particularly, the aviation regulatory authorities (the United States federal aviation administration) regulate the use of outdoor lasers in aircraft flight paths and prohibit visible lasers or high-powered non-visible lasers from being aimed at aircraft pilots, [20]. Hence, increasing the transmit power has a number of drawbacks. The packet size optimization was also proposed to enhance the data rate of the FSO, [21]. Though optimizing the packet size can guarantee the communication reliability, the overall rate cannot significantly increase because the amount of payload sent when the Signal to Noise Ratio (SNR) is low is lower than the ordinary packet size. On the other hand, introducing a technique that enhances the SNR provides a better data rate than decreasing the packet size when there is more signal deterioration.

To mitigate the atmospheric turbulences, a closed-form of mathematical expression is derived, [22], however, it cannot be applied to multiple objectives, for example, it doesn't discuss how to increase the range of communication. A self-healing Bessel beams accompanied by adaptive compensation techniques have proposed [23] which can reduce the inter-channel crosstalk and BER. It is good that the phase distortions caused by atmospheric turbulence are solved by integrating an adaptive compensation unit, however, as the optical coupler and polarizer can introduce additional processing complexity it cannot be suitable for low latency optical communications.

On the other hand, the performance of the FSO distance is evaluated for which the transmitted signal can be received without any error, [24]. Though the research has shown the understanding of the distance limit of the transmitted signal that can be received without (almost) error, the techniques to increase the distance without significantly affecting the error rate were not discussed.

Though fiber optic has tremendous advantages, for several years it was limited to the backhaul networks. The physical connection between the access network and the end-user was not benefited from the reliable and high bandwidth optical fiber nature. Finally, the FSO systems provide an innovative solution to this problem, however, the atmospheric challenges have continued as the main challenges to widely deploy the FSO, [25]. Hence, this research focuses on mitigating atmospheric problems.

Recently, OFDM-based radio over free-space optics have proposed and they have shown attractive link-distance enhancement on different atmospheric conditions, [26, 27]. However, both multiplexing to enhance the data rate and using low order modulation to enhance the reliability have the drawbacks of higher processing time and lower data rate, respectively, for future communications, 5G. Besides, the OFDM-based communication has peak-to-average power fluctuation problem, [28]. The network management technique was also proposed to enhance the communication performance, [29], however, the proposed BER and data rate are not achievable without the introduction of the FSO. The BER performance of wavelength selection has studied, [30], however, mitigating the atmospheric turbulences using the wavelength selection introduces operating spectrum inflexibility. To combat the FSO atmospheric turbulences, the robust modulation (BPSK) and spatial diversity techniques are used [31]. However, the robust modulation limits the throughput of the transmission and the spatial diversity introduces receiver computational complexity.

The coded-orthogonal frequency division multiple access (OFDM) has used to address the FSO atmospheric turbulences, [32], and of course, the coded and low-level modulation has shown better BER performance, however, this method of BER performance enhancement is well known. More specifically, increasing the reliability of the communication (BER reduction) at the expense of throughput reduction (because of the coding and decreasing the modulation order) is trivial engineering solution.

Hence, motivated by the attractive FSO characteristics an adaptive communication system is designed using an iterative optimization. The proposed technique adapts the atmospheric conditions including haze, fog, snow, mist, and rain conditions. More specifically, when the visibility of the transmission distance changes due to the atmospheric conditions, the amplifier works in a manner that covers the distance. On the other hand, the reliability of communication is guaranteed by the proposed iterative optimization technique. Quality of Service (QoS) is guaranteed by specifying the BER to a minimum possible value. The BER constancy is kept by the iterative optimization that maximizes the possible transmission distance without increasing the transmit power while still guaranteeing the QoS by minimizing the BER to an acceptable level.

## 2 System model

The FSO has attractive applications including better data rates, better security, cheap network installation, and license-free spectrum. Besides, it has better immunity from the electromagnetic interference because it cannot be detected using the RF meter, it is neither visible nor health hazardous, it can easily achieve very low BER, unlike RF antennas it doesn't have side lobes and the deployment is both cheap and quick. In contrary to their attractive benefits and applications, the FSO is often prone to atmospheric absorption, beam dispersion, rain, fog, snow, and shadows, [15–20].

Generally, the wireless optical link contains the transmitter, the atmospheric channel, and the receiver, Figure 1. The transmitter part transmits the signal in the wireless media by converting the electrical signal into the optical one using the optical modulator. Then, the optical signal propagates via the wireless medium and it is collected by the receiver and converted into a useful electrical signal. The transmitter subsystem consists of a pulse generator, line coder, modulator, optical power meter, spectrum analyzer, switching system, and optical amplifiers. The pulse generator generates pulses that carry the information in electrical form. The modulator converts the baseband signal into a high frequency that is suitable for transmission. The optical power meter measures the amount of power ready for transmission. The spectrum analyzer displays the input signal against the frequency. The fork and switching subsystems are used to select the path in which the circuit has to connect. This helps to decide whether to use one or more optical amplifiers or not. The switching decides based on the proximity of the receiver. More specifically, if the receiver is in closer proximity, then the amplification of the signal may not be required. However, when the receiver is far from the transmitter, then this switch connects to the amplifier and enables better signal quality. The optical amplifier increases the intensity of the signal which helps to easily fight the atmospheric effects. This enables better distance coverage without increasing the transmit power of the transmitter. The major signal attenuation happens in the wireless channel. This is because the atmospheric effects can significantly attenuate the signal. The overall attenuation is calculated as, [16]:

$$\alpha_{Total} = \alpha Fogy_\gamma + \alpha Snow_\gamma + \alpha Haze_\gamma + \alpha Rainy_\gamma + \alpha Mist_\gamma, \ dB/km$$

where, $\alpha$ is the attenuation and $\gamma$ is the operational wavelength in μm.

On the other hand, the receiver is comprised of optical amplifiers, photodetector, low pass filter, power meter, and BER analyzer. Similar to the optical amplifier used in the transmitter, it improves the received signal strength. The photodetector perceives the received optical signal and it converts it into electrical form. The low pass filter reduces the total environment noise by allowing to pass only a certain frequency of the signal. Finally, the BER analyzer determines the accuracy of the received signal. The BER averages the probability of correctly received bits out of the total transmitted bits.

$$BER = \frac{Number\ of\ errors}{Total\ number\ of\ bits\ sent}$$

On the other hand, the BER can be calculated from the SNR of the received information, [16]:

$$BER = \frac{2}{\pi \times SNR} \times e^{\left(\frac{-SNR}{8}\right)}$$

The output of the system is evaluated for three different optical transmission atmospheric conditions with the attenuation values 20 dB/km, 30 dB/km, and 70 dB/km using a BER analyzer.

The atmosphere is the gaseous layer that surrounds the planet. Fog is a thick cloud of tiny water droplets suspended in the atmosphere that restricts visibility. On the other hand, Smoke is a visible suspension of particles in the air, typically one emitted from a burning substance, for example, carbon. The Haze is another atmospheric phenomenon where the dust, smoke, and other dry particles make the sky unclear. Dust is a fine powder made up of very small pieces of earth or sand. The common atmospheric conditions and the corresponding signal attenuation values are summarized in Table 1.

In this paper, the optical amplifier operates without the need for conversion to electrical signals. Since the optical content of the signal is amplified, the SNR and hence the BER performs better (See Figure 2).

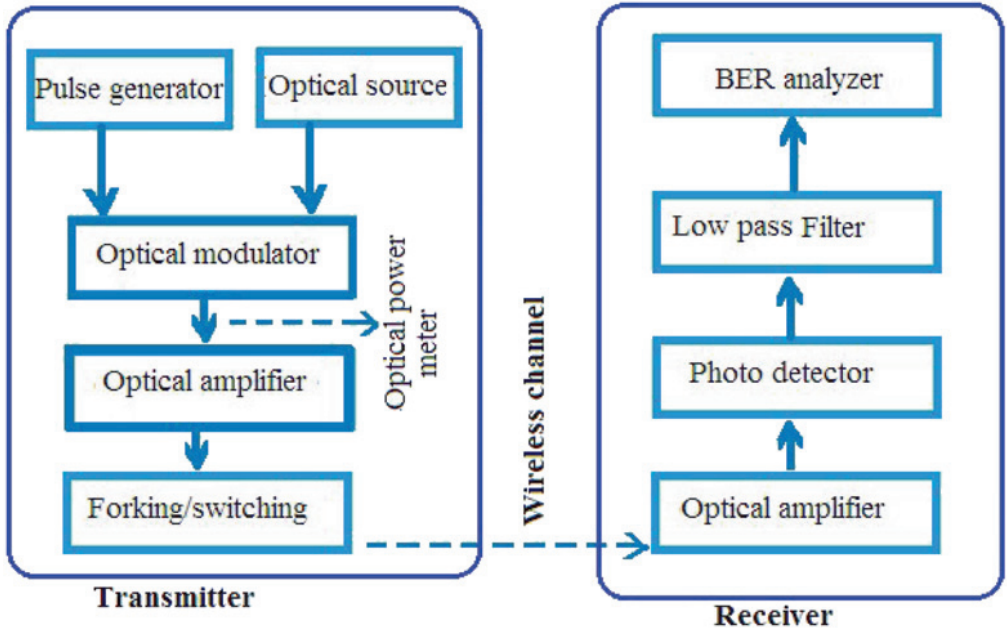

**Figure 1:** Overview of the optical wireless communication.

**Table 1:** FSO atmospheric conditions and the corresponding signal attenuation magnitudes.

| Atmospheric conditions | Attenuation (dB/km) |
|---|---|
| Haze | 10.94–20.68 |
| Rain | 6.0–30 |
| Mist | 28.56–31.45 |
| Snow | 40 |
| Fog | 70 |

The optical amplifier boosts the average power of the laser output and it also amplifies the weak signals before the photodetector detects the optical signal. This reduces the detection noise and hence decreases the BER of the communication. Especially in longer visibility distance optical communication, the optical power level should be raised before the information is lost in the noise. This is a big deal with wireless optical communication because the atmospheric conditions severely affect the signal. It is also well known that amplifiers do not only amplifies the amplitude or phase of the input signal but also introduces some noise. Hence, with this trade-off, the less the BER it results the better the significance of the amplifier.

As it is shown in Figure 3, the fork/switch circuit is used to adapt the distance of the receiver. If the communication distance is short enough, then the switch selects the circuit without the additional optical amplifier and it switches to the amplifier when the receiver is far from the transmitter. This adapting the visibility distance is especially important when the receiver and/or transmitter are mobile in nature.

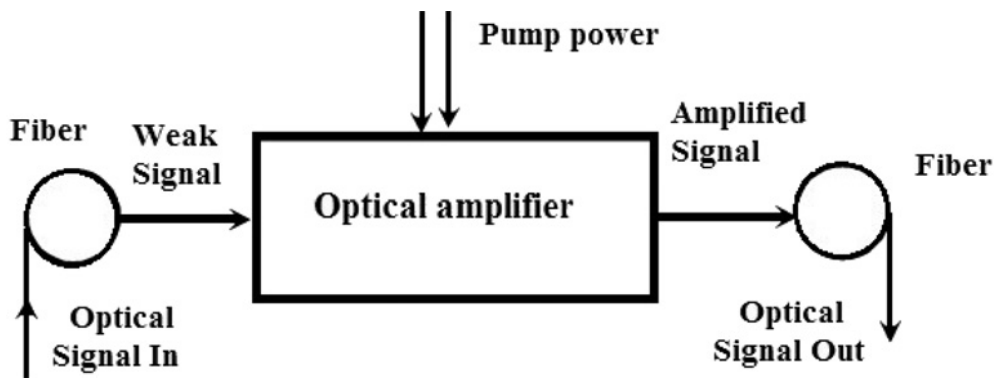

**Figure 2:** Overview of optical amplifier.

## 2.1 Optimization technique

The optimization technique is a mathematical procedure which applies a random starting test parameters to generate ordered improving approximate solutions for a certain problem. The optimization technique proposed in this paper is an iterative optimization that starts with termination criteria and constraint boundaries. The proposed iterative algorithm utilizes the problems whose solution fairly constrained by many service requirements and it avoids significant computational downsides.

The objective of this study is to maximize visibility distance and minimize error rate while the reliability and data rate are kept guaranteed. The proposed iterative optimization problem is to minimize the BER, given as a function $f$ of $N$ variables. The proposed iterative optimization is also going to maximize the visibility distance, given as a function $g$ of $M$ variables. Now let's find the minimizer, i.e. the point $x*$ such that

$$f(x*) \leq f(x), \quad \forall x \text{ near } x*$$

again, let's find the maximizer, i.e. the point $r^*$ such that

$$g(r*) \leq g(r), \quad \forall r \text{ near } r*$$

finally, let's express the overall problem as $\min_x f(x) \cup \max_r g(r)$, then we can have a combined function, $h$ with the $x$ and $r$ variables: $h(xr) = \frac{g(r)}{f(x)}$. Hence, the overall problem will be to maximize $h$:

$$\max_{xr} h(xr) = \frac{\max_r g(r)}{\min_x f(x)}$$

$$\text{Subjected}: \begin{cases} R \geq R_b \\ B \leq B_e \end{cases}$$

Where $R$ is the data rate that is constrained to be not less than $R_b$ and $B$ is the BER that is constrained to a minimum acceptable value, $B_e$, that satisfies the service requirement. This also guarantees the reliability of the optical communication.

## 3 Results and discussions

Using the proposed FSO, the end to end optical design is constructed as shown in Figure 3. The optical transmitter is fixed to a data rate of 10 Gbps. It is then encoded with the NRZ pulse generator. As an optical source, the CW laser is used and it is given 60 dBm of power and 1550 nm wavelength. With this in mind, the proposed visibility distance maximizing and BER minimizing mechanism have evaluated under different atmospheric conditions including Haze, rain, mist, and fog. The Q factor, BER, and received power are evaluated using OptiSystem 16.

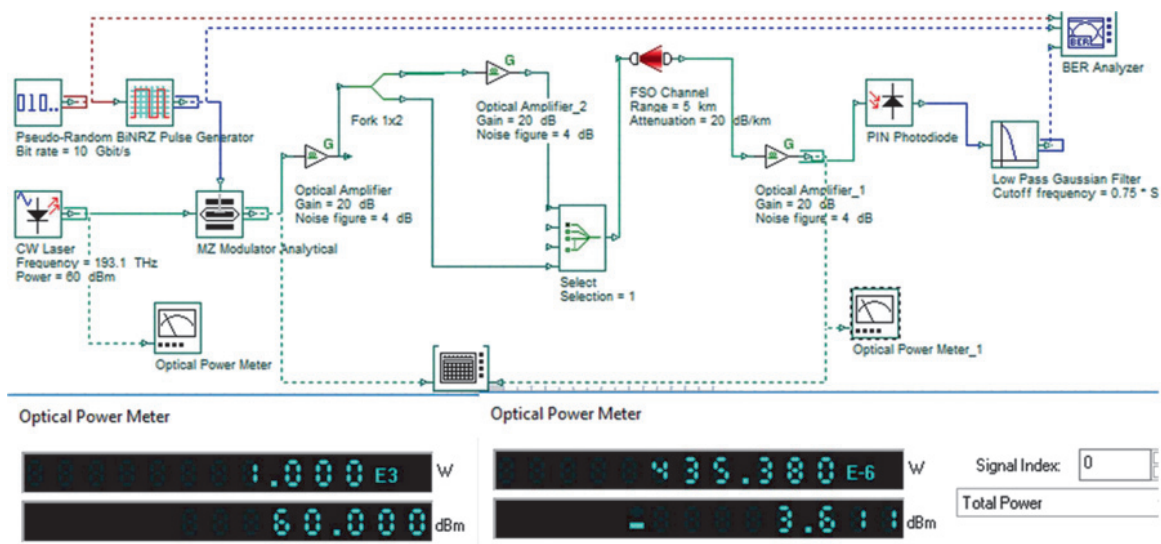

**Figure 3:** Schematic view of the adaptive FSO design.

## 3.1 Under haze atmospheric conditions

It is not surprising that the quality factor and signal power are reducing over the distance, shown in Figure 4(a) and (b). However, it is interesting that very long-distance propagation is achieved without increasing the transmit power of the transmitter. Since the optical signal has a lot of safety problems (including aviation exposure problems) and this effect increases when the transmit power increase. Hence, the proposed design is ideal as a reliable FSO communication is achieved without increasing the transmit power.

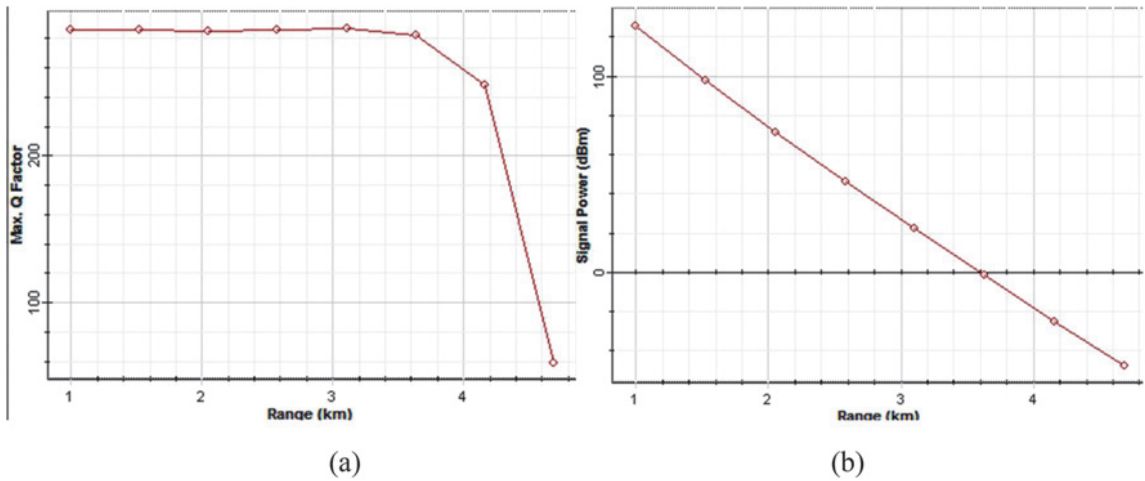

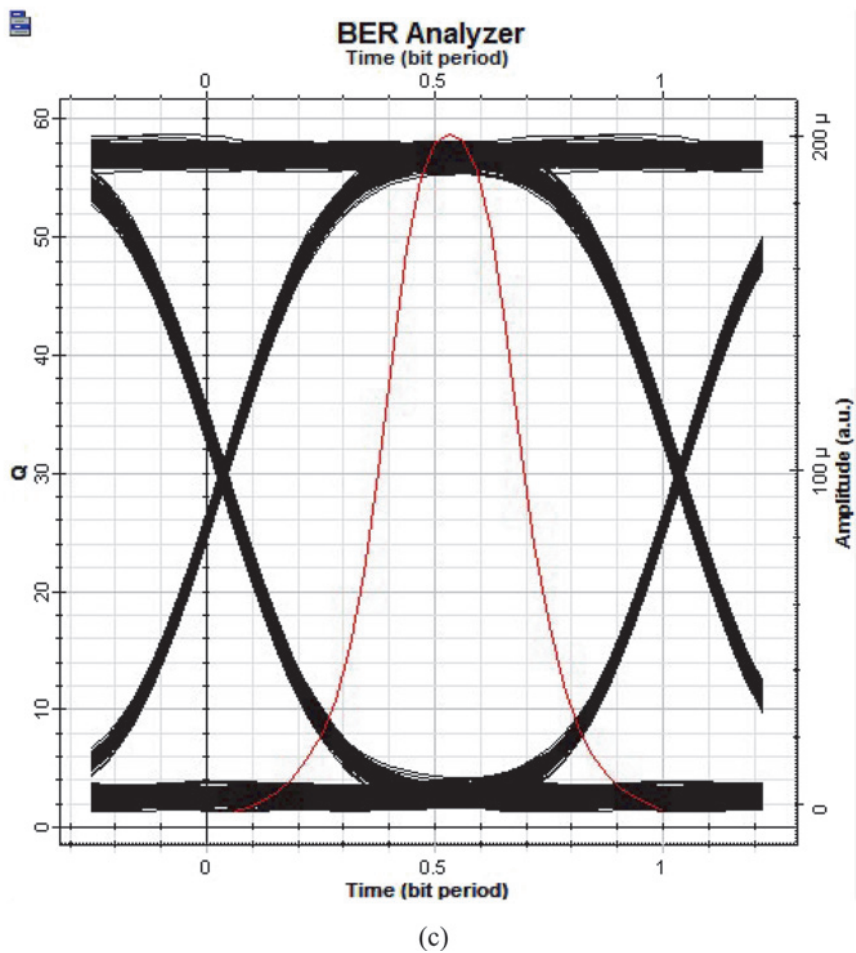

**Figure 4:** (a) Q factor, (b) received power and (c) eye pattern performances over distance.

Though the haze atmospheric wireless environment significantly attenuates the optical signal, the proposed adaptive and iterative algorithm has enabled 4685 m distance propagation without any repeaters. As can be shown in Figure 4(c), this long-distance optical signal propagation has achieved without spoiling the quality of service (i.e., the BER). The Q-factor measures the quality performance of the optical link. The longer the distance the more the Q-factor is decreased because the signal strength decreases with increasing link distance.

## 3.2 Under rain and mist atmospheric conditions

Similar to the Haze atmospheric wireless environment, the rain and mist have significant optical signal attenuation effects. Even the rain and mist have higher atmospheric effects than the haze atmospheric conditions. Though the atmospheric signal attenuation is increased, the optimal design has enabled us up to 3314 m distance propagation while still, the BER is almost zero, shown in Figure 5(a)–(c).

As the signal power decreases fast, as shown in Figure 5(b), the optical link is limited to a finite distance. Beyond a certain distance limit, the signal is no more useful as the atmospheric attenuation, scattering, and reflection erode the content of the signal. On the other hand, the noise overwhelms the signal at the receiver which results in difficult signal regeneration. Hence, the Q-factor decreases with distance, as shown in Figure 5(a).

Even at this adverse environment (rainy and misty), the new optical communication design has empowered us to communicate more than 3.3 km in an unlicensed spectrum. Unlike the many RF and optical communications, this remarkable distance coverage is not attained at the expense of throughput or at the expense of communication reliability. This unlicensed spectrum enables attractive distance coverage and it is also characterized 10 Gbps data rate and no error.

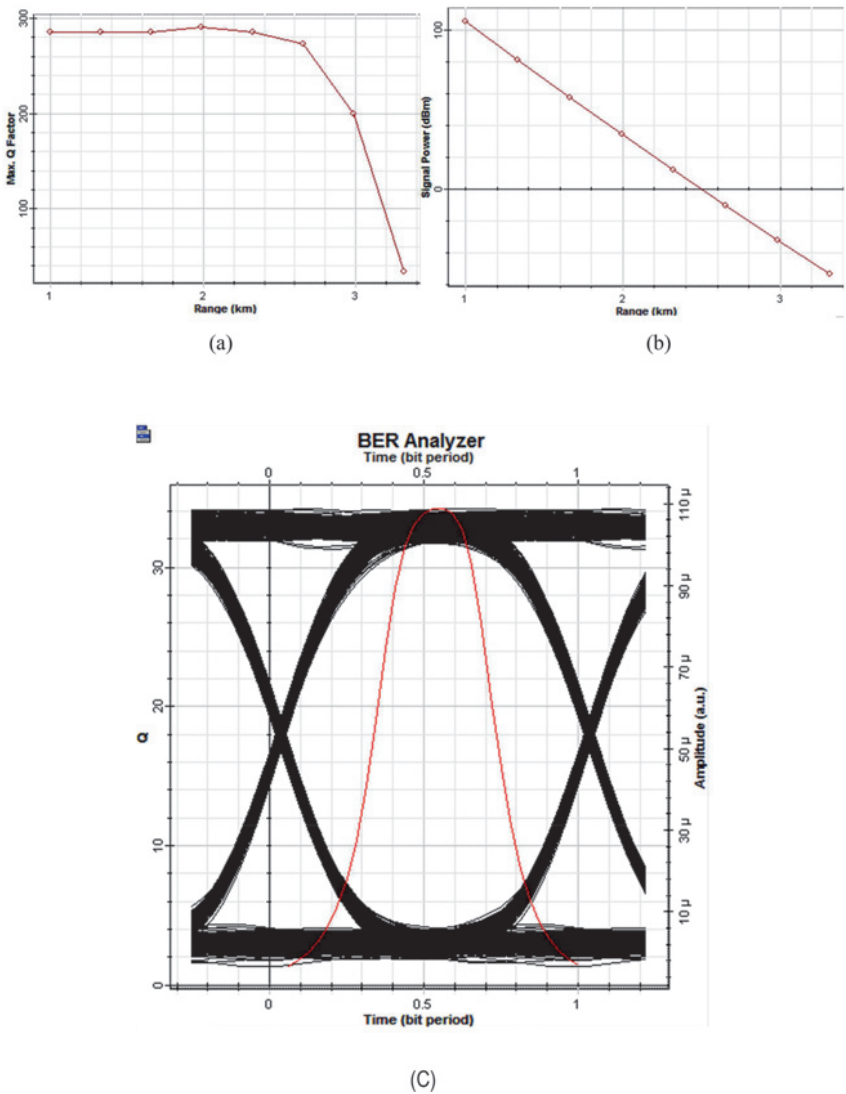

**Figure 5:** Q factor (a) & received power (b) performances under rain and mist conditions (c): Eye diagram under rainy and misty atmospheric conditions.

## 3.3 Under fog atmospheric conditions

The fogy atmospheric condition attenuates the optical signal much more than the rainy and misty atmospheric conditions (more than twice worse effect). This results in faster signal fall down, as shown in Figure 6(a) and (b). If the usual unlicensed optical wireless communication link is used, then the usual design doesn't enable longer distance communication. Hence, the optical communication design should adapt its parameters and circuitry to enable reliable communication when the atmospheric conditions are changing. The maximum distance possible for reliable communication has reached 1550 m with the proposed optical design. Though reliable communication is possible beyond this distance, it may not be achieved at a low cost. For example, it is well known that further distance coverage will be achieved at the expense of increased transmit power. However, the increase in transmit power contradicts the energy efficiency use case of the 5G communication networks and it introduces safety problems.

The Q-factor of the rainy and misty is still fine while the 1550 m distance is covered, Figure 6(c). Furthermore, the BER is negligible at this adverse condition.

Though the fogy atmospheric condition significantly attenuates the FSO signal (70 dB/km), with the help of the proposed iterative optimization and new design, more than 1.5 km distance FSO communication is possible without the need for additional repeaters. It is also good to notice that the proposed design adapts the atmospheric conditions while collecting channel state conditions.

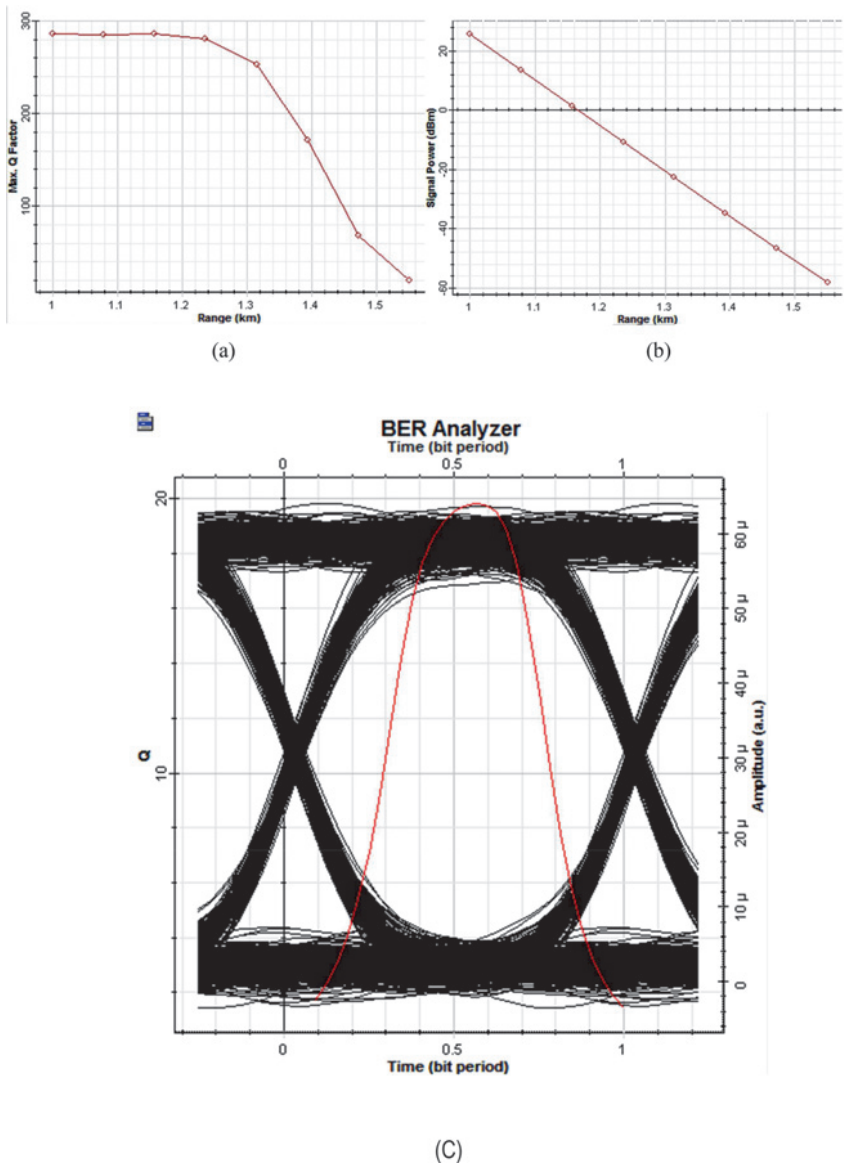

Figure 6: Q factor (a) & received power (b) performances under rainy and misty conditions (c): eye diagram under rainy and misty atmospheric conditions.

The literature [17] has increased the visibility distance by applying three concatenated couplers. However, though it has used more optical devices, the distance increment was just a small amount. In this paper, the optimization of the amplifier length has increased the visibility distance while the QoS and data rate are not deteriorated by the increase in distance.

As we can observe in Table 2 and Figure 7, there are different visibility distance enhancement mechanisms. From the literature, we can clearly observe that different operating bands ($S$, $C$, $L$) have different atmospheric effect resistances. Different filters and different modulation technics also result in different atmospheric effect tolerance and hence different visibility distances. However, the visibility distance enhancements were only limited incremental effects. The proposed solution has optimized to maximize the distance while the throughput and QoS are guaranteed. While still, the Q factor is better than the literature, significant visibility distance enhancement is achieved.

Tables 2–4 briefly present the performance enhancement achieved in this paper compared to the literature. Mainly, the proposed work has conducted to enhance the visibility distance of wireless optical communication. However, increasing the visibility distance at the expense of the BER, data rate, and Q-factor is not sounding scientific design. Hence, the BER, data rate, and Q-factor are made better than or equal to the literature. In all cases, the proposed work has overperformed to the existing works because of the optimization and modified optical link designs.

Figures 8 and 9 have shown performance enhancement in terms of the visibility distance. In all the rainy, misty and foggy, the proposed work has achieved longer distance coverage without affecting the signal quality. The Figures can also indicate that the threshold for reliable QoS communication is limited to these maximum distances according to the proposed design. More specifically, when the atmospheric condition is known to be in the hazy region, then the proposed link budget enables 4685 m distance without deteriorating the signal quality. A similar, conclusion holds for the misty, rainy, and foggy atmospheric conditions.

To sum up, the proposed transceiver design enables better communication performance in the FSO scope. Both the iterative optimization and adaptive characteristics of the transceiver made the design handy for the FSO in any atmospheric environment. For several years, the significance of FSO was limited because of the atmospheric challenges. Of course, if the atmospheric problems are solved or reduced, the extremely low BER, high bandwidth, and license-free long-distance communication are possible. The optical fiber was also very expensive and it requires a lot of deployment costs. However, the FSO is quite simple in deployment while still an affordable cost. Due to its incapability to the front wireless communication, the fiber optics communication was limited to the backhaul networks before the introduction of the FSO. Hence, the FSO is an attractive solution to the alarmingly increasing traffic demand for 5G.

**Table 2:** Summary results under Haze atmospheric conditions.

| Reference | Attenuation (dB/km) | Visibility distance (m) | Q factor | BER | Data rate (Gbps) |
|---|---|---|---|---|---|
| [13], 2016 (ASK & PPM) | 20 | 100 | 3.86 | 5.03e$^{-5}$ | 10 |
| [14], 2015 | 20 | 800 | 21.08 | – | 2.5 |
| [15], 2017 ($S$, $C$ & $L$ bands) | 20 | 500 | 45.96 | 0 | 10 |
| [17], 2018 (ASK & PSK) | 20 | 3500 | 57.0486 | 0 | 10 |
| Proposed (Optimization) | 20 | 4685 | 58.762 | 0 | 10 |

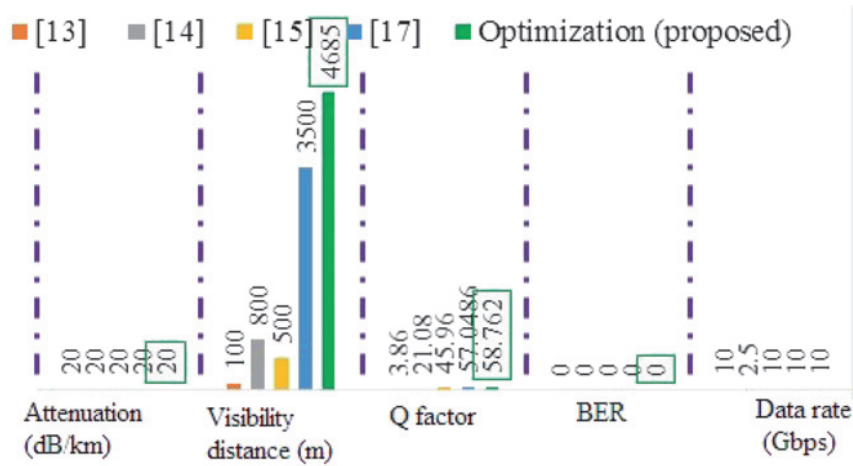

**Figure 7:** Visibility distance comparison under Hazy atmospheric conditions.

Table 3: Summary results under rainy and misty atmospheric conditions.

| Reference | Attenuation (dB/km) | Visibility distance (m) | Q factor | BER | Data rate (Gbps) |
|---|---|---|---|---|---|
| [13], 2016 (ASK & PPM) | 30 | 100 | 3.86079 | $5.025e^{-5}$ | 2.5 |
| [14], 2015 | 30 | 800 | 9.62 | – | 2.5 |
| [15], 2017 (S, C & L bands) | 30 | 500 | 45.96 | 0 | 10 |
| [17], 2018 (ASK & PSK) | 30 | 2500 | 34.0397 | $2.855e^{-254}$ | 10 |
| Proposed (Optimization) | 30 | 3314 | 34.20 | $1.1544e^{-256}$ | 10 |

Table 4: Summary results under Fogy atmospheric conditions.

| Reference | Attenuation (dB/km) | Visibility distance (m) | Q factor | BER | Data rate (Gbps) |
|---|---|---|---|---|---|
| [15], 2017 (S, C & L bands) | 70 | 500 | 2.934 | 0.00160863 | 10 |
| [16], 2016 (850, 1310 & 1550 nm windows) | 70 | – | 2.59 | 0.00153508 | 2.5 |
| [17], 2018 (ASK & PSK) | 70 | 1200 | 17.5354 | $3.841e^{-69}$ | 10 |
| Proposed (Optimization) | 70 | 1550 | 19.8265 | $8.72786e^{-88}$ | 10 |

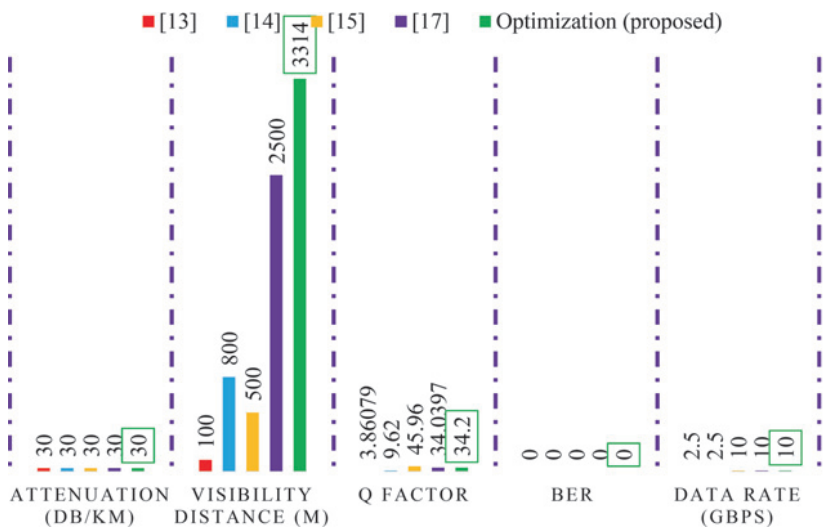

Figure 8: Visibility distance comparison under rainy & misty atmospheric conditions.

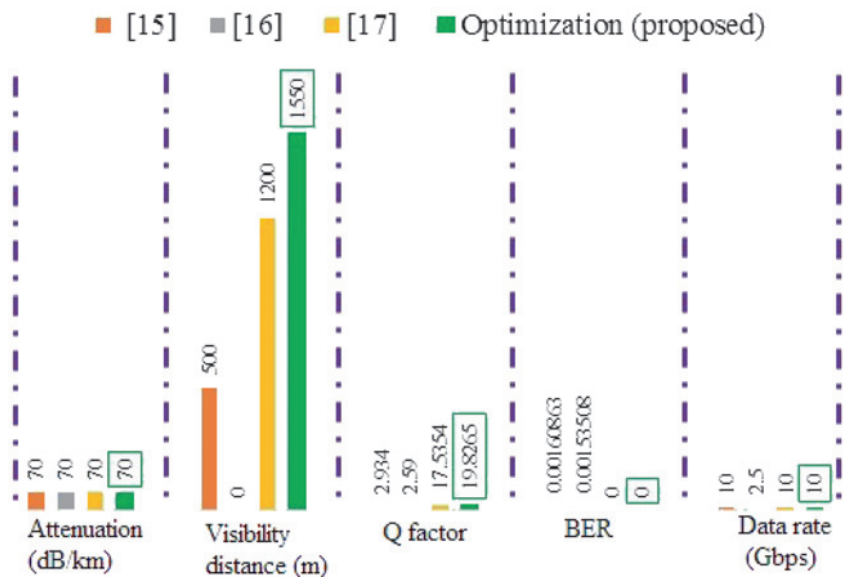

Figure 9: Visibility distance comparison under fogy atmospheric conditions.

## 4 Conclusion

Even though the FSO enables attractive communication characteristics, the wireless link easily deteriorates the optical signal. The data rate and BER are highly weather conditions dependent which severely attenuate the signal. To decrease the atmospheric conditions, this study has proposed an iterative optimization algorithm and an enhanced optical communication link design.

The optimized optical link design is evaluated its performance in terms of visibility distance, quality factor, BER, and Eye diagram at hazy, misty, rainy, and foggy atmospheric conditions. The performance evaluation has conducted while the QoS is guaranteed using the reliability and data rate. Keeping the BER, data rate, and Q-factor are greater than or equal to the recent researches, the visibility distance is maximized by advancing the optical link design and by optimizing the optical amplifier length. For any given atmospheric condition, the maximum possible guaranteed QoS visibility distance is determined and adapting the atmospheric condition is possible using the newly proposed design. In general, the simulation results have shown that better visibility distance, Q factor, and less BER are achieved at the expense of little system complexity.

**Author contribution:** All the authors have accepted responsibility for the entire content of this submitted manuscript and approved submission.
**Research funding:** The article didn't receive any fund.
**Conflict of interest statement:** The authors declare no conflicts of interest regarding this article.